\makeatletter
\newcommand{\dontusepackage}[2][]{%
  \@namedef{ver@#2.sty}{9999/12/31}%
  \@namedef{opt@#2.sty}{#1}}
\makeatother
\dontusepackage{subfigure}

\documentclass[]{article}

\usepackage{lmodern}
\usepackage{amssymb,amsmath}
\usepackage{ifxetex,ifluatex}
\usepackage[usenames,dvipsnames]{color}
\usepackage{fixltx2e} 
\ifnum 0\ifxetex 1\fi\ifluatex 1\fi=0 
  \usepackage[T1]{fontenc}
  \usepackage[utf8]{inputenc}
\else 
  \ifxetex
    \usepackage{mathspec}
    \usepackage{xltxtra,xunicode}
  \else
    \usepackage{fontspec}
  \fi
  \defaultfontfeatures{Mapping=tex-text,Scale=MatchLowercase}
  
\fi
\IfFileExists{upquote.sty}{\usepackage{upquote}}{}
\IfFileExists{microtype.sty}{%
\usepackage{microtype}
\UseMicrotypeSet[protrusion]{basicmath} 
}{}
\usepackage[margin=1.0in,bottom=1.5in]{geometry}
\usepackage[square,numbers,sort&compress]{natbib}
\usepackage{listings}
\usepackage{graphicx}
\makeatletter
\def\maxwidth{\ifdim\Gin@nat@width>\linewidth\linewidth\else\Gin@nat@width\fi}
\def\maxheight{\ifdim\Gin@nat@height>\textheight\textheight\else\Gin@nat@height\fi}
\makeatother
\setkeys{Gin}{width=\maxwidth,height=\maxheight,keepaspectratio}
\usepackage{caption}
\usepackage{float}
\usepackage{subfig}
\ifxetex
  \usepackage[setpagesize=false, 
              unicode=false, 
              xetex]{hyperref}
\else
  \usepackage[unicode=true]{hyperref}
\fi
\hypersetup{breaklinks=true,
            bookmarks=true,
            pdfauthor={},
            pdftitle={Uncertainty quantification in imaging and automatic horizon tracking---a Bayesian deep-prior based approach},
            colorlinks=true,
            citecolor=black,
            urlcolor=blue,
            linkcolor=black,
            pdfborder={0 0 0}}
\usepackage[preprint]{neurips_2019}
\usepackage{url}
\usepackage{booktabs}
\usepackage{amsfonts}
\usepackage{nicefrac}
\usepackage{microtype}

\title{Uncertainty quantification in imaging and automatic horizon tracking---a
Bayesian deep-prior based approach}
\author{Ali Siahkoohi, Gabrio Rizzuti, and Felix J. Herrmann\\School of
Computational Science and Engineering,\\Georgia Institute of
Technology\\\texttt{\{alisk,\phantom{\ }rizzuti.gabrio,\phantom{\ }felix.herrmann\}@gatech.edu}}
\date{}

\begin{document}
\maketitle
\begin{abstract}
In inverse problems, uncertainty quantification (UQ) deals with a
probabilistic description of the solution nonuniqueness and data noise
sensitivity. Setting seismic imaging into a Bayesian framework allows
for a principled way of studying uncertainty by solving for the model
posterior distribution. Imaging, however, typically constitutes only the
first stage of a sequential workflow, and UQ becomes even more relevant
when applied to subsequent tasks that are highly sensitive to the
inversion outcome. In this paper, we focus on how UQ trickles down to
horizon tracking for the determination of stratigraphic models and
investigate its sensitivity with respect to the imaging result. As such,
the main contribution of this work consists in a data-guided approach to
horizon tracking uncertainty analysis. This work is fundamentally based
on a special reparameterization of reflectivity, known as ``deep
prior''. Feasible models are restricted to the output of a convolutional
neural network with a fixed input, while weights and biases are Gaussian
random variables. Given a deep prior model, the network parameters are
sampled from the posterior distribution via a Markov chain Monte Carlo
method, from which the conditional mean and point-wise standard
deviation of the inferred reflectivities are approximated. For each
sample of the posterior distribution, a reflectivity is generated, and
the horizons are tracked automatically. In this way, uncertainty on
model parameters naturally translates to horizon tracking. As part of
the validation for the proposed approach, we verified that the estimated
confidence intervals for the horizon tracking coincide with geologically
complex regions, such as faults.
\end{abstract}

\section{Introduction}\label{introduction}

Aside from well data, seismic images are used to delineate stratigraphy.
Automatic tracking of the horizons from seismic images
\citep{wu2018least, peters2019} is becoming a more broadly adapted
technology to determine the stratigraphy. While this is a beneficial
development, questions remain on its reliability. Even though it is
clear that the accuracy of horizon tracking is directly linked to the
quality of the seismic image, principled investigations on the risks
associated with identifying these horizons are often lacking. Clearly,
failure to include uncertainty on tracked horizons may have major
implications on the identifications of risks. For this purpose, we
propose a technique where we directly translate uncertainty in the image
to uncertainty in the tracked horizons. We do this by drawing samples
from the posterior distribution of the image followed by automatic
horizon tracking. We gain access to samples from the posterior by
casting seismic imaging as a Bayesian inversion problem, regularized
with a convolutional neural network (CNN), and running preconditioned
stochastic gradient Langevin dynamics
\citep[pSGLD,][]{li2016preconditioned, welling2011bayesian}, a Markov
chain Monte Carlo (MCMC) sampler. pSGLD generates an ensemble of
reflectivity models, each of which are likely solutions to the imaging
problem. Aside from providing information on the point-wise standard
deviation of the reflectivity, we use these samples to calculate
confidence intervals for our automatically tracked seismic horizons.
Compared to conventional imaging and manual tracking of horizons, our
combined approach of generating samples from the image posterior and
automatic horizon tracking allows us to assess risk in a systematic
manner.

While some progress has been made in Uncertainty Quantification (UQ) for
seismic imaging
\citep{doi:10.1046/j.1365-246X.2002.01847.x, malinverno2004expanded, malinverno2006two, MartinMcMC2012, Ely2018, fang2018uqfip, zhu2018seismic, izzatullah2019bayesian, visser2019bayesian, zhao2019gradient},
further advances are hampered by computational challenges and by over
simplifying and often biasing assumptions on parameterizations of the
prior and posterior distributions. Motivated by recent developments in
machine learning and geophysics
\citep{Lempitsky, Cheng_2019_CVPR, gadelha2019shape, liu2019deep, wu2019parametric, shi2020deep, siahkoohi2020EAGEdlb},
we overcome these challenges by parameterizing the unknown reflectivity
model in terms of a randomly initialized CNN, where its weights are
Gaussian random variables, and sampling the the posterior.
\citet{siahkoohi2020EAGEdlb} showed that CNNs act as a regularizer
because CNNs tend to generate naturally looking images as long as we
prevent them from overfitting. Contrary to early work on deep prior
where the optimization is stopped early to prevent overfitting,
\citet{siahkoohi2020EAGEdlb} demonstrated that pSGLD avoids fitting the
noise and provides samples from the posterior. Through these samples, we
are not limited to the maximum a posteriori estimator (MAP), and we are
able to compute the conditional mean (known to be relatively more robust
with respect to noise artifacts than MAP estimations), and point-wise
standard deviation of the image. We track the horizons automatically for
each sample of the posterior for the image. This allows us to add
confidence intervals to the tracked horizons, which now reflect
uncertainties in the image. While we use a deterministic automatic
horizon tracking approach \citep{wu2018least}, our method will be able
to accommodate stochastic horizon tracking schemes---e.g., horizons
tracked by many interpretors \citep{schaaf2019quantification} or by
different event tracking softwares.

There have been numerous efforts to incorporate ideas from deep learning
in seismic processing and inversion
\citep{ovcharenko2019deep, rizzuti2019EAGElis, siahkoohi2019dlwr, siahkoohi2019transfer, siahkoohi2019srmedl, zhang2019regularized}
but there have been relatively few attempts towards UQ.
\citet{mosser2018stochastic} use a pretrained generative CNN as a prior
in seismic waveform inversion and sample the posterior by running a
variant of stochastic gradient Langevin dynamics
\citep[SGLD,][]{welling2011bayesian} on the latent variable of the
generative CNN. To handle situations where there is no access to
training pairs, \citet{herrmann2019NIPSliwcuc} introduces a formulation
that combines handcrafted priors with deep priors to jointly solve the
inverse problem and train a model capable of directly sampling the
posterior. In this work, we also aim to come up with an unsupervised
approach to UQ, but now take it a step further to include a principled
way to assess the risk of conducting an additional task on the image,
namely tracking the reflector horizons. To our knowledge, our approach
is close to recent work by \citet{adler2018task}, who proposed a
Bayesian framework for jointly performing inversions and tasks. However,
our approach differs because it explicitly uses samples from the
posterior on the images, by tying uncertainties in the imaging to
uncertainties during the subsequent task of of horizon tracking. Our
approach also differs fundamentally from other recently developed
automatic seismic horizon trackers based on machine learning \citep[see
e.g.][]{peters2019} because horizon uncertainty is ultimately driven by
data (through the intermediate imaging distribution), and not from label
(control point) uncertainty.

Our work is organized as follows. We first mathematically formulate how
to sample from the image's posterior distribution by introducing the
likelihood function and prior distribution involving the deep prior.
Next, we present our approach to quantify the uncertainty in horizon
tracking, using samples from the posterior distribution for the image
using pSGLD. We conclude by showcasing our approach on synthetic example
derived from a 2D portion of a real migrated image of the
\href{https://wiki.seg.org/wiki/Parihaka-3D}{Parihaka-3D} dataset
\citep{Veritas2005, WesternGeco2012}, which is used to evaluate
automatic horizon tracking algorithms.

\section{Bayesian Seismic Imaging}\label{bayesian-seismic-imaging}

The objective of seismic imaging is to estimate the reflectivity model,
denoted by $\delta \mathbf{m}$, given observed data,
$\delta \mathbf{d}_{i}$, a smooth background squared-slowness model,
$\mathbf{m}_0$, and assumed to be known seismic source signatures,
$\mathbf{q}_i$, where $i = 1,2, \cdots , N$ and $N$ is the number of
shot records. To formulate a posterior distribution, in addition to a
prior distribution on the reflectivity, we need to specify a likelihood
function, $p_{\text{like}}$ \citep{tarantola2005inverse}. Assuming
zero-mean Gaussian noise, the negative log-likelihood of the observed
data ($\{\delta \mathbf{d}_i\}_{i=1}^{N}$) can be written as follows:

\clearpage
\begin{equation}
\begin{aligned}
&\ - \log p_{\text{like}} \left ( \left \{ \delta \mathbf{d}_{i}\right \}_{i=1}^N \normalsize{|} \delta \mathbf{m} \right )   = -\sum_{i=1}^N  \log p_{\text{like}} \left ( \delta \mathbf{d}_{i}\normalsize{|} \delta \mathbf{m} \right ) \\
&\ = \frac{1}{2 \sigma^2} \sum_{i=1}^N  \| \delta \mathbf{d}_i- \mathbf{J}(\mathbf{m}_0, \mathbf{q}_i)  \delta \mathbf{m}\|_2^2 \quad + \underbrace {\text{const}}_{\text{Ind. of }   \delta \mathbf{m}}. \\
\end{aligned}
\label{imaging-likelihood}
\end{equation}
 In this equation, $\sigma^2$ is the variance of noise in the data and
$\mathbf{J}$ the linearized Born scattering operator computed for a
known smooth background model ($\mathbf{m}_0$) and source signatures.

Because of noisy data and a nullspace of the Born scattering operator,
maximization of the likelihood does not lead to acceptable images and we
need to add prior information on the image as a regularization. While
many choices for selecting prior distributions exist, they tend to bias
MAP or other estimates. We circumvent this bias by using randomly
initialized CNNs as priors for the reflectivity model, an approach
recently advocated in the literature
\citep{Lempitsky, Cheng_2019_CVPR, gadelha2019shape, liu2019deep, wu2019parametric, shi2020deep, siahkoohi2020EAGEdlb}.
To this end, we reparameterize the unknown reflectivity model by a
randomly initialized CNN---i.e.,
$\delta \mathbf{m} = g(\mathbf{z}, \mathbf{w})$, where the vector
$\mathbf{z} \sim \mathrm{N}( \mathbf{0}, \mathbf{I})$ is the fixed input
to the CNN. The vector $\mathbf{w}$ represents the unknown CNN weights.
In this formulation, the prior is made of a combination of the
functional form of the CNN and a Gaussian prior on the weights---i.e.,
$\mathbf{w} \sim \mathrm{N}(\mathbf{0}, \frac{1}{\lambda^2}\mathbf{I})$,
where $\lambda$ is a hyperparameter. Based on these definitions, the
negative log-posterior becomes
\begin{equation}
\begin{aligned}
&\ - \log p_{\text{post}} \left ( \mathbf{w} \normalsize{|}  \left \{ \delta \mathbf{d}_{i}\right \}_{i=1}^N \right ) \\
&\ =  -  \left [ \sum_{i=1}^{N} \log p_{\text{like}} \left ( \delta \mathbf{d}_{i}\normalsize{|}\mathbf{w}  \right ) \right ]  - \log p_{\text{prior}} \left ( \mathbf{w} \right ) + \underbrace {\text{const}}_{\text{Ind. of } \mathbf{w}} \\
&\ = \frac{1}{2 \sigma^2} \sum_{i=1}^N  \| \delta \mathbf{d}_i- \mathbf{J}(\mathbf{m}_0, \mathbf{q}_i) {g} (\mathbf{z}, \mathbf{w}) \|_2^2 + \frac{\lambda^2}{2} \| \mathbf{w} \|_2^2 +   \text{const,} \\
\end{aligned}
\label{imaging-obj}
\end{equation}
 where $p_{\text{post}}$ and $p_{\text{prior}}$ are the posterior and
prior distributions, respectively. In the next section, we present how
to draw samples from this posterior distribution.

\section{UQ for Seismic Imaging}\label{uq-for-seismic-imaging}

MCMC sampling is a well established method to draw samples from
unnormalized probability density functions, which in principle makes a
suitable candidate for sampling the posterior defined in
Equation~\ref{imaging-obj}. Unfortunately, MCMC methods become
computationally expensive for high-dimensional problems
\citep{fang2018uqfip, izzatullah2019bayesian, zhao2019gradient}. We
employ several strategies to mitigate the costs. For instance, we work
with few shot gather stacks randomly formed at each iteration. Secondly,
we use a preconditioned version of SGLD proposed by
\citet{li2016preconditioned}. After inclusion of the adaptive diagonal
preconditioning matrix $\mathbf{M}_k$ at the $k^{\text{th}}$ iteration
(see \citet{li2016preconditioned} for for detail) the pSGLD update reads
\begin{equation}
\begin{aligned}
&\ \mathbf{w}_{k+1} = \mathbf{w}_{k} - \frac{\epsilon}{2} \mathbf{M}_{k} \nabla_{\mathbf{w}} L^{(i)}(\mathbf{w}_{k}) + \boldsymbol{\eta}_k, \quad \boldsymbol{\eta}_k \sim \mathrm{N}( \mathbf{0}, \epsilon \mathbf{M}_{k}), \\
\end{aligned}
\label{sgld}
\end{equation}
 where
$L^{(i)} (\mathbf{w}) = \frac{N}{2 \sigma^2} \| \delta \mathbf{d}_i- \mathbf{J}(\mathbf{m}_0, \mathbf{q}_i) {g} (\mathbf{z}, \mathbf{w}) \|_2^2 + \frac{\lambda^2}{2} \| \mathbf{w} \|_2^2$.
In this expression, we evaluate the likelihood in
Equation~\ref{imaging-obj} by only using the randomly selected
$i^{\text{th}}$ term in the sum. The parameter $\epsilon$ is the step
size. Under certain technical conditions, including properly decaying
step sizes, and after an initial ``burn-in'' phase, the above iterates
correspond to samples from the posterior distribution
\citep{li2016preconditioned}. With these samples, we can approximate
expectations, such as the conditional mean, via a sample average over
$T$ realizations---i.e., we have
\begin{equation}
\begin{aligned}
\widehat { \delta \mathbf{m}} &\  = \mathbb{E}_{\mathbf{w} \sim p_{\text{post}} ( \mathbf{w} \normalsize{|}  \left \{ \delta \mathbf{d}_{i}\right \}_{i=1}^N )} \left [{g}( \mathbf{z}, \mathbf{w}) \right ]  \simeq \frac{1}{T}\sum_{j=1}^{T} {g} \left  ( \mathbf{z}, \widehat{\mathbf{w}}_j \right ), \\
\end{aligned}
\label{integration}
\end{equation}
 where
$\widehat{\mathbf{w}}_j \sim p_{\text{post}} ( \mathbf{w} \normalsize{|} \left \{ \delta \mathbf{d}_{i}\right \}_{i=1}^N ), \ j=1, \ldots, T$.
In a similar fashion, point-wise standard deviation can be computed and
since it quantifies variability amongst the samples, it contains useful
UQ information.

\section{Seismic horizon tracking with
UQ}\label{seismic-horizon-tracking-with-uq}

While having access to UQ information in the form of estimated
point-wise standard deviation can be useful, one is often more
interested in how these uncertainties derived from the data propagate
into risks associated with certain tasks conducted on the image. For
this purpose, we consider deterministic seismic horizon tracking, which
we denote by the mapping $\mathcal{H}$. Conceptually, this nonlinear
mapping represents deterministic actions of automatic horizon trackers
or of reliable, consistent human interpreters. In both cases, horizon
tracking is not informed by the data other than that is provided with a
migrated image. Without loss in generality, we use the automated horizon
tracking approach introduced by \citet{wu2018least}, which uses local
slopes of the imaged reflectivity to track horizons seeded by user
specified control points.

To compute uncertainties on the tracked horizons, we pass the obtained
samples from the posterior distribution for the image to the automatic
horizon tracking software. Next, by Monte-Carlo integration, we
approximate the first and (point-wise) second moment of the posterior
distribution for the horizons. The deterministic automatic horizon
tracking can be replaced by a nondeterministic approach, reflecting
stochasticity in the horizon tracking. Contrary to most existing
automatic horizon trackers, the uncertainty in the tracked horizons
presented here is due to noise in the shot records and not due
uncertainties in the control points as is more commonly studied
\citep[see e.g.][]{peters2019}.

\section{Implementation}\label{implementation}

The gradient computations required by Equation~\ref{sgld} involve
actions of the linearized Born scattering operator and its adjoint and
the gradient of the CNN with respect to its weights. For maximal
performance, we use Devito \citep{devito-compiler, devito-api} for the
wave-equation based simulations and we integrate matrix-free
implementation of these operators into PyTorch \citep{NEURIPS2019_9015}.
In this way, we are able to compute the gradients required in
Equation~\ref{sgld} with automatic differentiation. For the CNN
architecture, we follow \citet{Lempitsky}. We use the automated horizon
tracking \href{https://github.com/xinwucwp/mhe}{software} introduced by
\citet{wu2018least}. For more details on our implementation, please
refer to our code on
\href{https://github.com/slimgroup/Software.SEG2020/tree/master/siahkoohi2020SEGuqi}{GitHub}.

\section{Numerical experiments}\label{numerical-experiments}

\begin{figure}
\centering
\subfloat[\label{imaging-dm}]{\includegraphics[width=0.500\hsize]{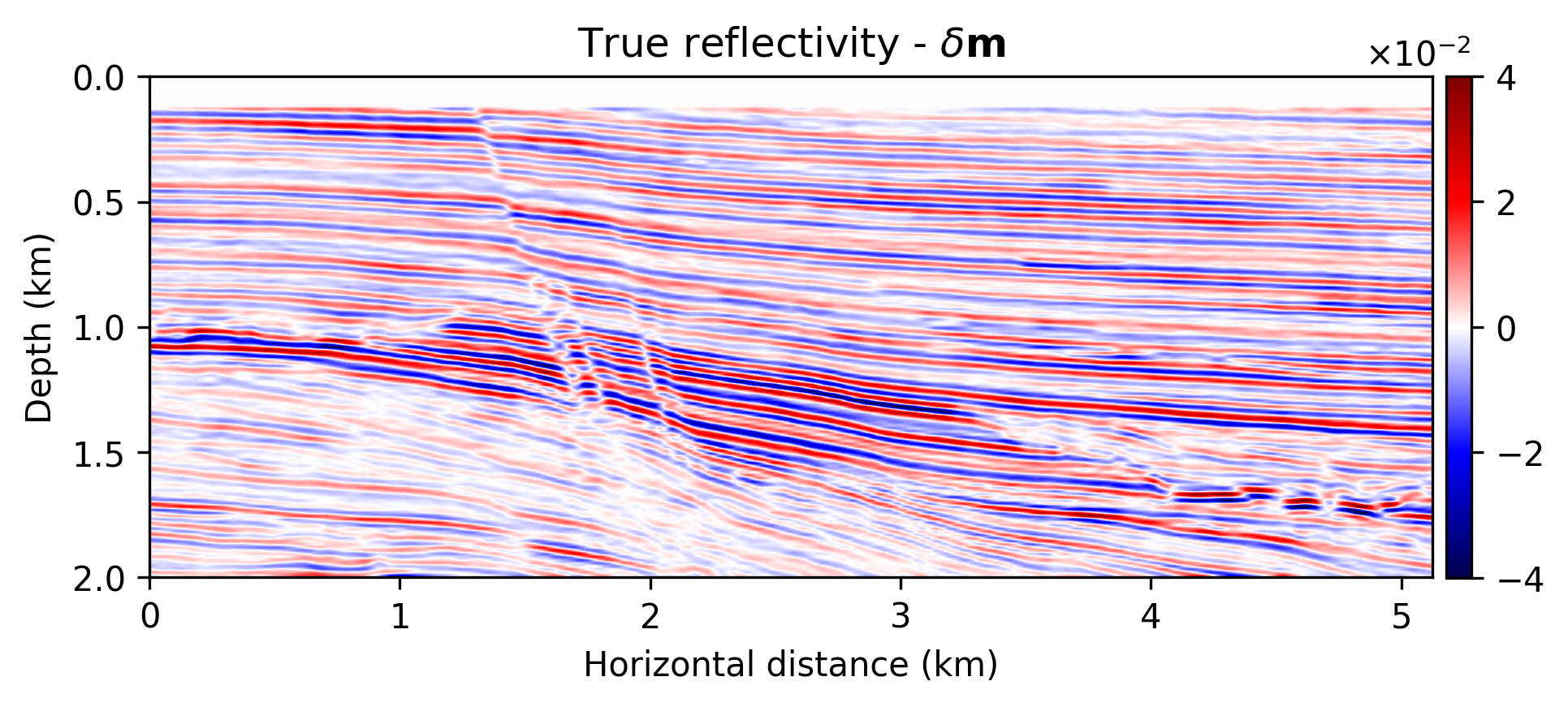}}
\subfloat[\label{imaging-mle}]{\includegraphics[width=0.500\hsize]{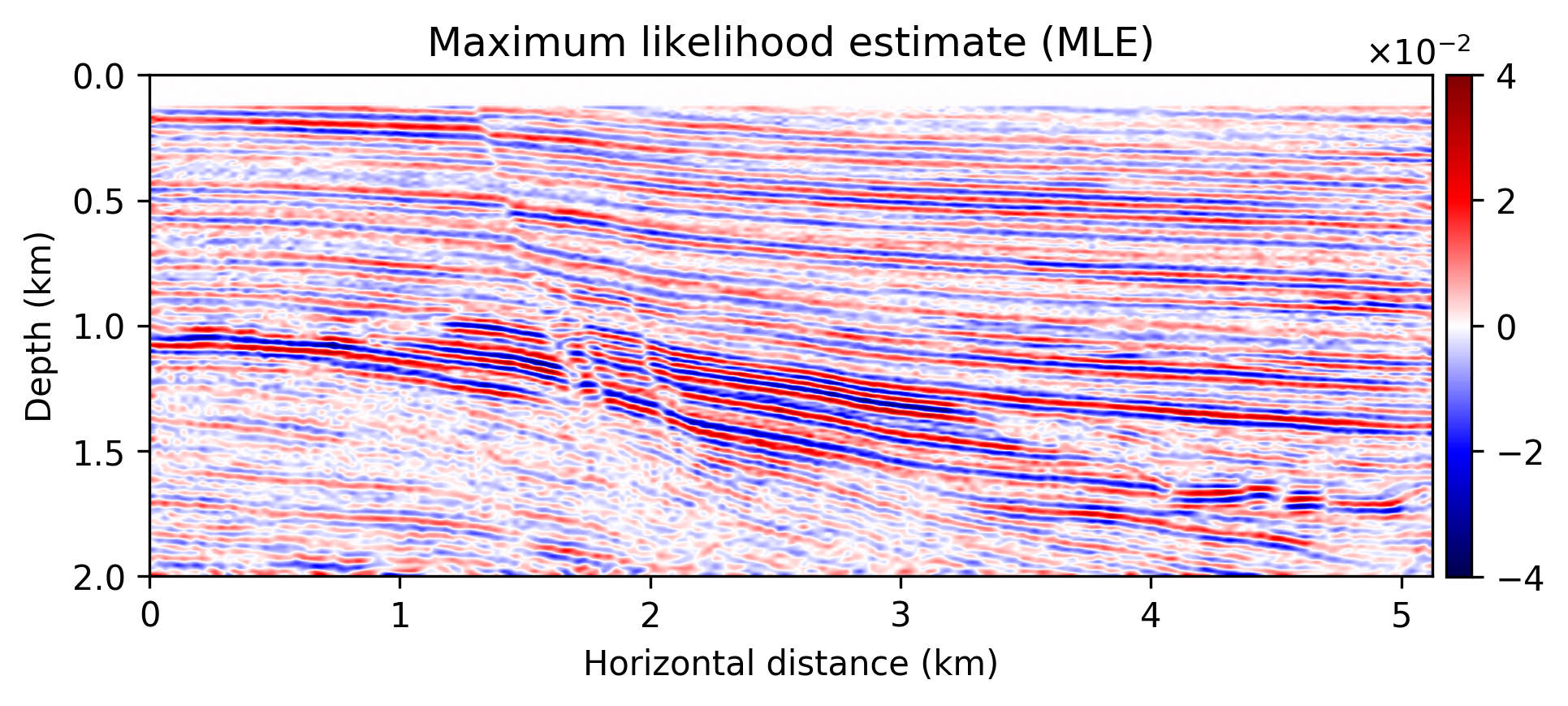}}
\\
\subfloat[\label{imaging-mean}]{\includegraphics[width=0.500\hsize]{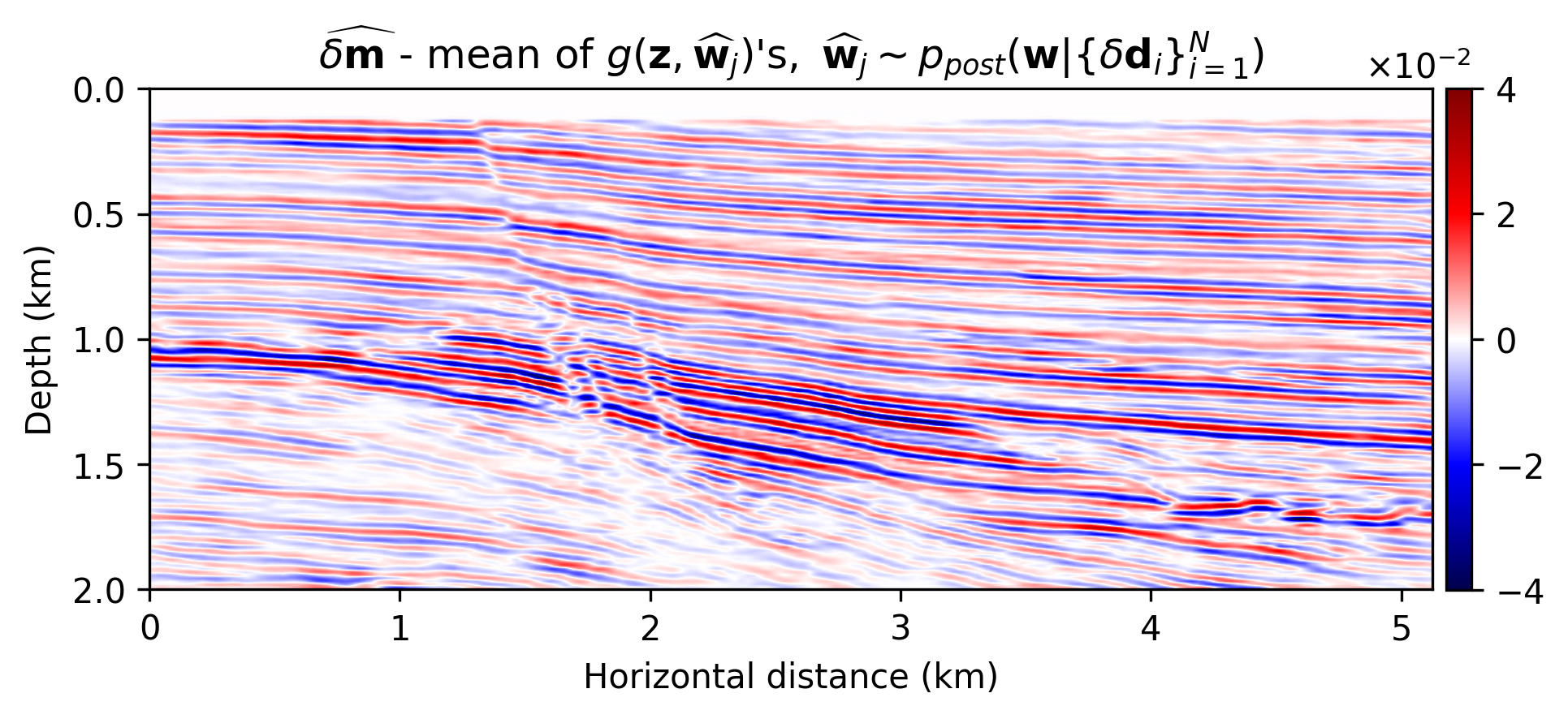}}
\subfloat[\label{imaging-std}]{\includegraphics[width=0.500\hsize]{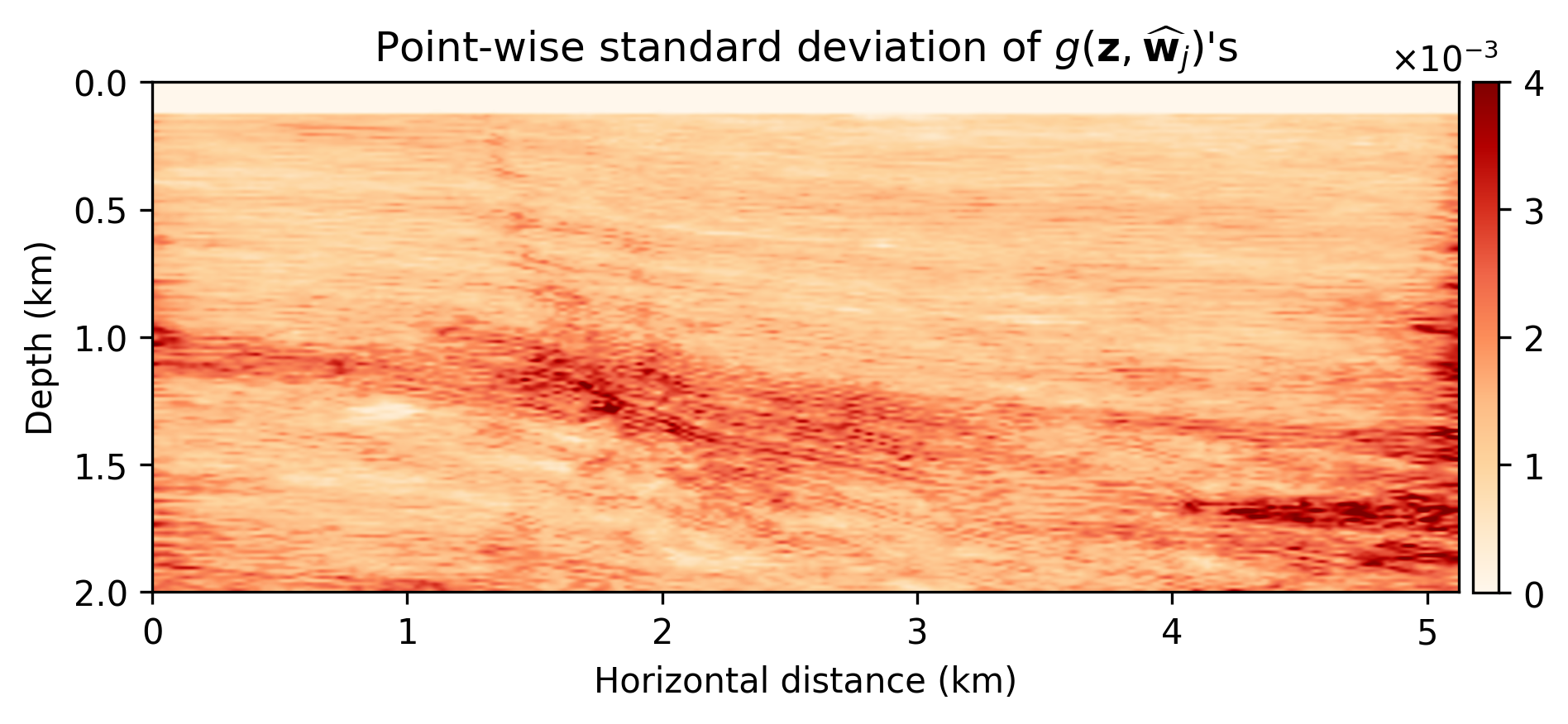}}
\caption{UQ in imaging. a) True model. b) Maximum-likelihood estimate.
c) Conditional mean, $\widehat { \delta \mathbf{m}}$. d) The point-wise
standard deviation among samples drawn from the
posterior.}\label{imaging-results}
\end{figure}

To demonstrate the performance of our approach, we consider a ``quasi''
real field data example that derives from a 2D subset of the real
Kirchoff time migrated
\href{https://wiki.seg.org/wiki/Parihaka-3D}{Parihaka-3D} dataset
released by the New Zealand government and used to test the seismic
horizon tracker developed by \citet{wu2018least}. We call our experiment
quasi real because we generate synthetic data from this ``true'' imaged
reflectivity (see Figure~\ref{imaging-dm}) using our linearized Born
scattering operator for a made up, but realistic, smoothly varying
background model $\mathbf{m}_0$. To ensure good coverage, we simulate
$205$ shot records sampled with a source spacing of $25\, \mathrm{m}$.
Each shot is recorded over $1.5$ seconds with $410$ fixed receivers
sampled at $12.5 \mathrm{m}$ spread across survey area. The source is a
Ricker wavelet with a central frequency of $30\,\mathrm{Hz}$. To mimic a
more realistic imaging scenario, we add a significant amount of noise to
the shot records, yielding a low signal-to-noise ratio of the
``observed'' data of $-18.01\, \mathrm{dB}$. To limit the computational
costs (= number of wave-equation solves), we work with a single
simultaneous source, made of a Gaussian weighted source aggregate, per
gradient calculation in Equation~\ref{sgld}. After an initial burn in of
$3\mathrm{k}$ iterations (about $15$ passes over the data), we sample
every $20^{\text{th}}$ iteration of the pSGLD iterations in
Equation~\ref{sgld}. After extensive parameter testing, we set the step
size $\epsilon=0.002$ and penalty parameter $\lambda^2=200$. The
$\sigma^2=0.01$ equals the variance of the noise we added to the shot
records.

After running $10\mathrm{k}$ iterations (about $49$ passes through the
data), we sampled $T=351$ realizations from the posterior distribution
on the image. To demonstrate the effect of regularization with the deep
prior, we first compute the maximum-likelihood estimate by minimizing
Equation~\ref{imaging-likelihood}. To avoid overfitting we stop early.
The result of this exercise is included in Figure~\ref{imaging-mle}. As
expected this result is noisy and above all lacks crucial details and
continuity. The estimate for the conditional mean,
$\widehat{\delta \mathbf{m}}$, on the other hand, is much improved,
clean and, contains many of the details present in the original ``true''
reflectivity (cf.~Figures~\ref{imaging-dm} --~\ref{imaging-mean}).
However, the samples of the posterior do show considerable (up to 10\%)
variability, as observed in Figure~\ref{imaging-std}, where the
point-wise standard deviation is, as expected, large in areas of complex
geology (e.g.~near faults and tortuous reflectors) and in areas with a
relatively poor illumination.

\begin{figure}
\centering
\subfloat[\label{horizon-CM}]{\includegraphics[width=0.468\hsize]{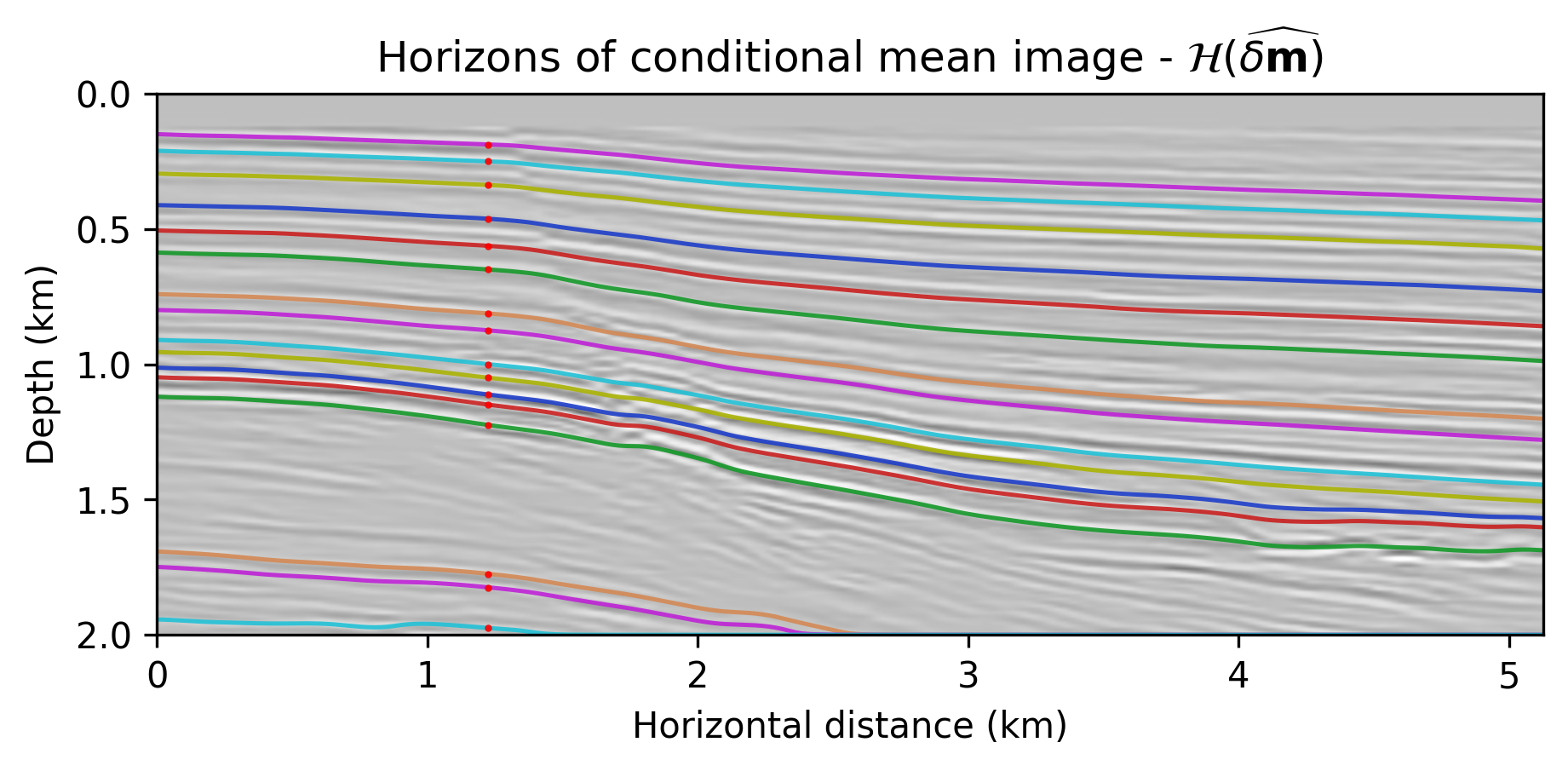}}
\subfloat[\label{horizon-49}]{\includegraphics[width=0.468\hsize]{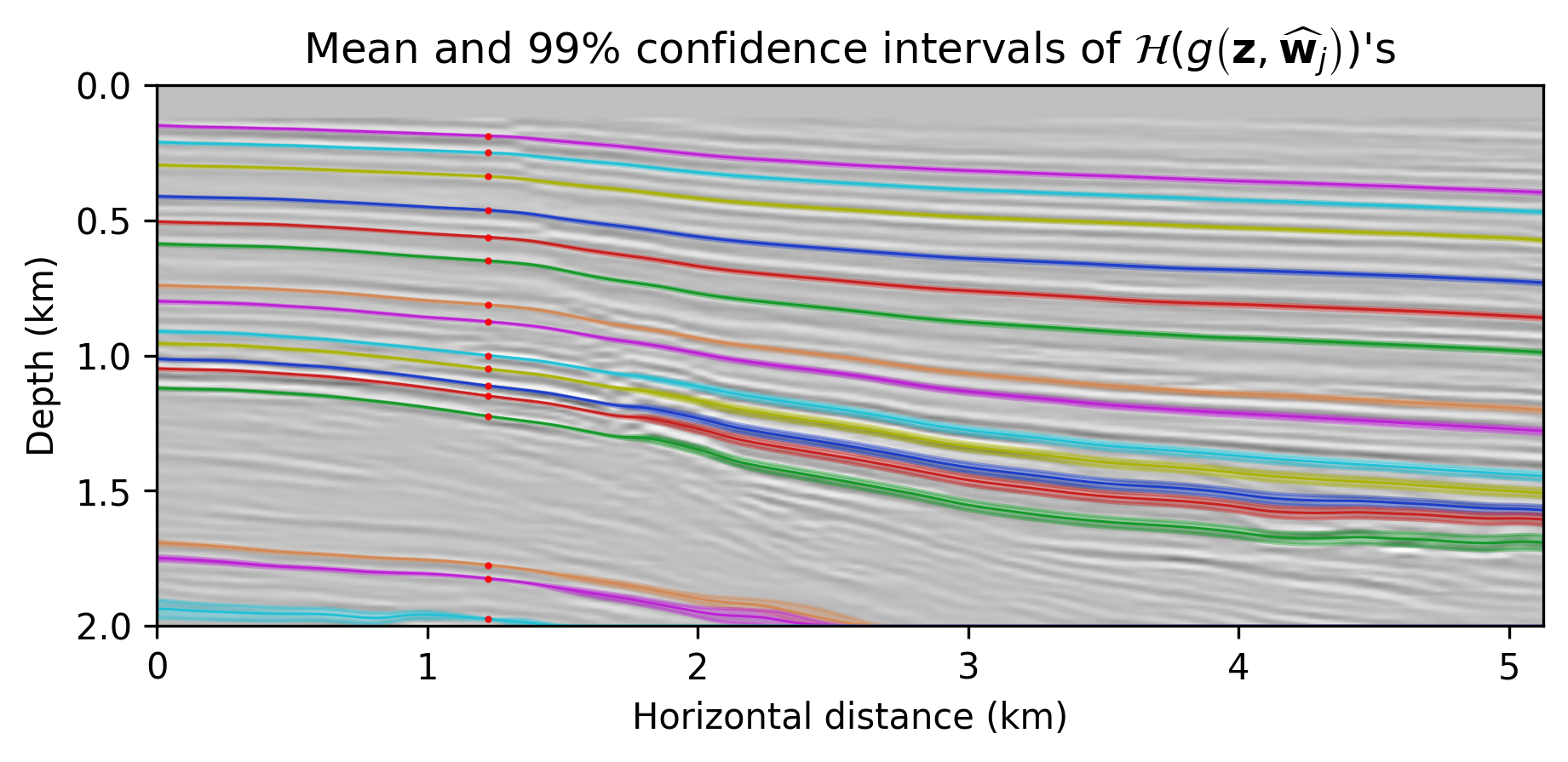}}
\\
\subfloat[\label{horizon-120}]{\includegraphics[width=0.468\hsize]{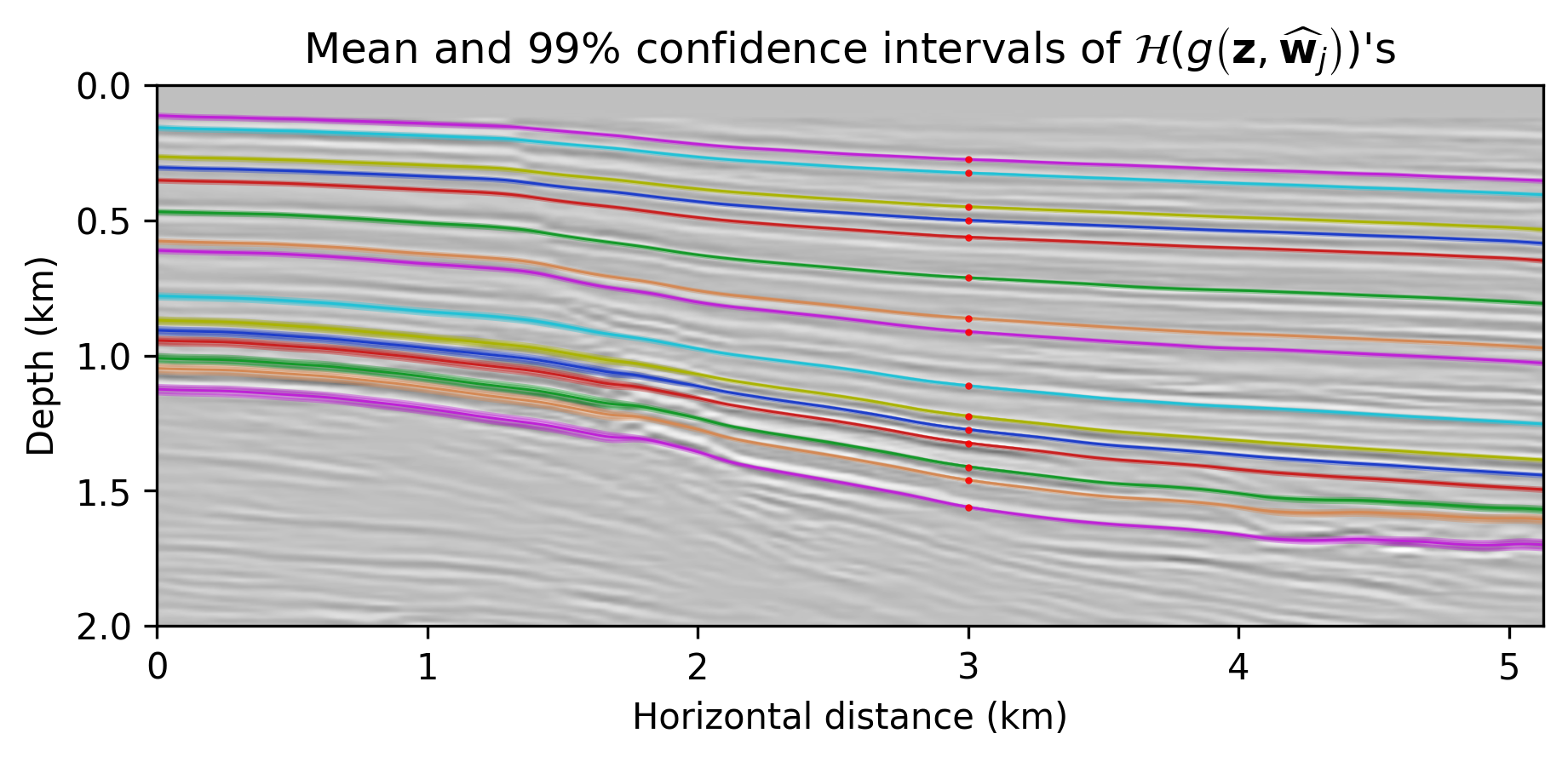}}
\subfloat[\label{zoom-49}]{\includegraphics[width=0.234\hsize]{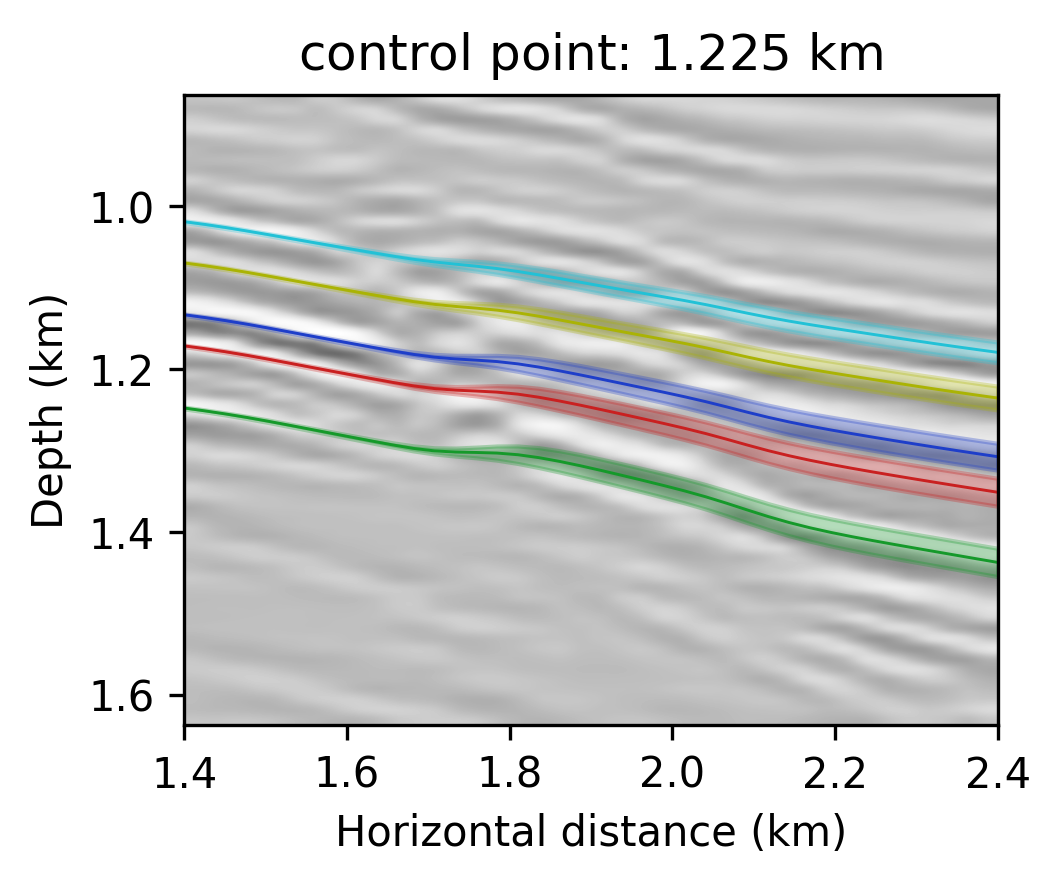}}
\subfloat[\label{zoom-120}]{\includegraphics[width=0.234\hsize]{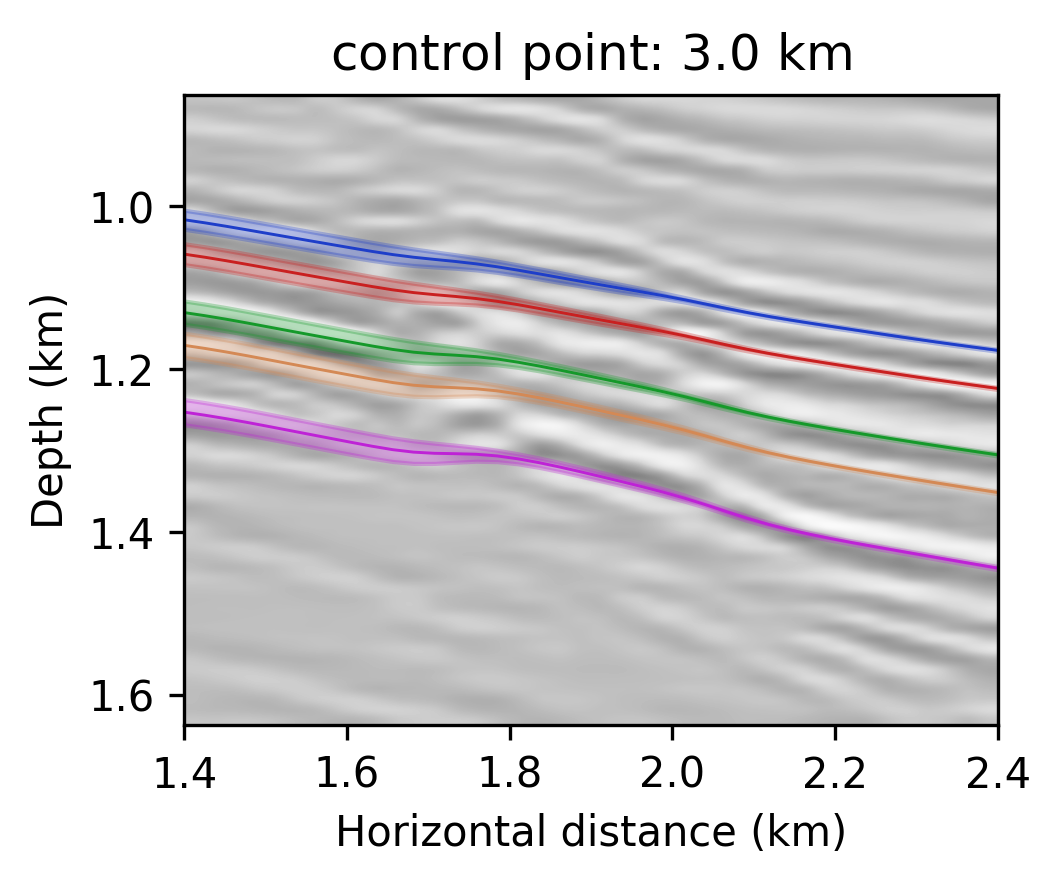}}
\caption{UQ in horizon tracking. a) Horizons of the conditional mean. b,
c) Mean and confidence intervals of posterior for horizons with control
points at $1.225$ $\mathrm{km}$ and $3$ $\mathrm{km}$, respectively. d,
e) Figures~\ref{horizon-49} and~\ref{horizon-120} restricted to a region
with faults.}\label{horizon-results}
\end{figure}

To illustrate how information on the posterior can be used to assess
uncertainties in horizon tracking, we use the proposed Monte-Carlo
sampling procedure to estimate the conditional mean and the associated
$99$\% confidence interval for a number of reflector horizons. To guide
us, we first track horizons on the conditional mean estimate---i.e.,
$\mathcal{H}( \widehat{\delta \mathbf{m}})$. We used this estimate to
select control points, the red dots located at $1.225$ $\mathrm{km}$
horizontal location, that seed the horizon tracker in both directions.
While these tracked horizons are close to the conditional mean,
$\mathbb{E}_{\mathbf{w} \sim p_{\text{post}} ( \mathbf{w} \normalsize{|} \left \{ \delta \mathbf{d}_{i} \right \}_{i=1}^N )} \left [ \mathcal{H} \left ( {g}(\mathbf{z},{\mathbf{w}} \right )) \right ]$,
their associated confidence intervals in shaded colors exhibit
considerable variation. We observe this behavior for different locations
of the control points (cf.~Figures~\ref{horizon-49}
and~\ref{horizon-120}) and as we move across faults, areas tortuous
reflectivity and into areas of lessened illumination near the edges and
in the deeper parts of the image. Depending on the location of the
control points, the uncertainty, as expected, increases at the opposite
side of the fault. This increase in the size of the confidence interval
is due the increased variability amongst samples of the posterior in the
region of increased complexity near the fault something we clearly
observe in the zoomed figures included in Figures~\ref{zoom-49}
and~\ref{zoom-120}.

\section{Conclusions}\label{conclusions}

Because of the high dimensionality and complexity of seismic imaging
problems, uncertainty quantification remains extremely challenging. In
this work, we present a solution for 2D problems by parameterizing
seismic images in terms of convolutional neural networks (CNNs) and
optimization of the network weights via preconditioned stochastic
gradient Langevin dynamics. We showed that the functional form of CNNs,
combined with a Gaussian prior on its weights, act as a regularizer and
partially circumvent imaging artifacts. Access to the posterior
distribution allows us to compute conditional mean, point-wise standard
deviation, and confidence intervals for automatically tracked reflector
horizons. This is accomplished in a sequential fashion, by computing
uncertainties on the imaging result via Monte-Carlo sampling first, and
then pushing forward those uncertainties on the tracked horizons.
Contrary to most existing horizon tracking approaches, our confidence
intervals are driven by noise in the shot records and not from control
point errors. In this work, control points were calculated from the
conditional mean imaging result, but information coming from well logs
can be in principle integrated as well. Our approach can further include
control point uncertainty, or applied to other tasks, such as image
segmentation. This work is, to our knowledge, one of the first instances
where data errors---e.g., due to noise or linearization approximations,
are systematically mapped to confidence interval of some imaging result
attributes.

\section{Acknowledgments}\label{acknowledgments}

The authors thank Charles Jones for constructive conversations and
comments.

\bibliography{abstract}

\end{document}